\documentclass[manuscript]{aastex}
\newcommand\curl{\nabla \times}

\usepackage{epstopdf}

\begin{document}

\title{Consequences of Magnetic Field Structure for Heat Transport in Magnetohydrodynamics}
\author{Shule Li\\ Adam Frank and Eric Blackman}
\affil{Department of Physics and Astronomy, University of Rochester, Rochester, NY, 14627}
\email{shuleli@pas.rochester.edu}
%\maketitle

\begin{abstract}
Interfaces  between hot and cold magnetized plasmas exist in various astrophysical contexts, for example 
where hot outflows impinge on an ambient interstellar medium (ISM). It is of interest to understand how 
the structure of the magnetic field spanning the interface  affects the temporal evolution of the temperature 
gradient. Here we explore the relation between the magnetic field topology and the heat transfer rate by 
adding various fractions of  tangled vs. ordered field across a hot-cold interface allow the system to evolve 
to a steady state. We find a simple  mathematical relation for the rate of heat conduction as a function of 
the initial ratio of ordered to tangled field across the interface. We discuss potential implications for the 
astrophysical context of magnetized wind blown bubbles (WBB) around evolved stars.
\end{abstract}


\section{Introduction}
Interfaces between hot and cold plasmas can occur in astrophysics where understanding the rate of thermal conduction 
may be an important part of  the the astrophysical phenomenology. One example  occurs in  wind blown bubbles (WBB) of 
evolved stars where magnetized hot supersonic outflow shock heats the cooler ambient magnetized  interstellar medium.  
For such WBB, there  are examples where  the presumed shock heated bubble is cooler than expected if only radiative cooling 
is considered (\citet{zhe11}). A possible explanation is that heat loss through the interface of hot bubble into the 
cold shell via thermal conduction reduces the temperature of the hot bubble (\citet{zhe98}, \citet{zhe00}).  
However the source of heat into the cold side of the interface will continuously evaporate material there and potentially 
induce interface instabilities and mass mixing (\citet{sto09}) that could tangle the magnetic field. Understanding 
the thermal conduction  and its dependence on magnetic structure is important for determining the thermal properties of the plasma 
on either side of the interface.
  
A second example is the unexpected slow mass deposition rate of the cooling flows in some  galaxy cores which might be inhibited 
by a restricted thermal conduction (\citet{ros89}, \citet{bal08}, \citet{mik11}). In the  intracluster 
medium (ICM), the tangled magnetic field can potentially produce a strongly anistropic thermal conductivity that may significantly 
influence  temperature and density profiles (\citet{cha04}; \citet{mar04}; \citet{nar01}, \citet{mik11}). 

For the ISM and ICM, it is usually valid to assume that the electrons are totally inhibited from moving across field lines 
(\citet{mcc11}), as the  electron mean free path is much greater than the electron gyroradius. The magnetic field structure 
therefore plays a key role in controlling the rate of thermal conduction since electrons can  move freely  only along the field lines. 
The result is a strong thermal conductivity parallel to the field lines and a weak conductivity across the field lines.

The quantitative subtleties  of how a complicated  magnetic field structure affects thermal conduction for raises the open question 
of whether there is a simple measure of field tangling that allows a practical but reasonably accurate correction to the isotropic 
conduction coefficient  for arbitrarily tangled fields. In this context, two classes of problems can be distinguished. The first is 
the conduction in a medium for which  forced velocity flows drive  turbulence, which in turn  tangles the field into a statistically 
steady state turbulent spectrum  (\citet{tri89}, \citet{tao95}; \citet{mar04}). The second is the case in which the flow is laminar and 
the level of conduction inhibition is compared when the field starts from initial states of different levels of tangling subject to an 
imposed temperature difference across an interface. This second problem is the focus of our preset paper.
 
Using the ASTROBEAR  magnetohydrodynamics code  with anisotropic thermal conduction, we investigate the influence of initial magnetic 
structure on thermal conduction in an otherwise laminar flow. The key questions  we address  are: (1) does the interface become unstable? 
(2) how fast is the thermal conduction across the interface compared to the unmagnetized case? 

We study these questions using different initial magnetic  configurations imposed on a planar thermal interface to determine how the 
conduction depends on the amount of field tangling across the interface.

In section 2, we review the basic equations of MHD with anisotropic thermal conduction. In sections 3 and 4 we provide detailed description 
of the simualtion setup. In section 5 and 6 we present the simulation results and analyses. In section 7, we discuss the simulation results 
in the context of the   WBB cooling problem and the cooling flow problem in cores of galaxy clusters. The appendix provides more detailed 
information on the testing of the ASTROBEAR code.

\section{MHD Equations with Anisotropic Heat Conduction}
The MHD equations with anisotropic heat conduction that we will solve are given by:
\begin{equation}
\frac{\partial \rho}{\partial t}+\nabla \cdot (\rho \bf{v}) = 0,
\end{equation}
\begin{equation}
\frac{\partial (\rho \textbf{v})}{\partial t}+\nabla \cdot [\rho \textbf{vv}+(p+\frac{B^2}{8\pi})\textbf{I}-\frac{\textbf{BB}}{4\pi}] = 0,
\end{equation}
\begin{equation}
\frac{\partial \textbf{B}}{\partial t}+\nabla \times (\textbf{v}\times \textbf{B}) = 0,
\end{equation}
\begin{equation}
\frac{\partial E}{\partial t}+\nabla \cdot [\textbf{v}(E+p+\frac{B^2}{8\pi})-\frac{\textbf{B}(\textbf{B}\cdot \textbf{v})}{8\pi}]+\nabla \cdot Q = 0,
\end{equation}
where $\rho$, $\textbf{v}$, $\textbf{B}$ and $p$ are the density, velocity, magnetic field, and pressure, and  
 $E$ denotes the total energy given by
\begin{equation}
E = \epsilon+p\frac{\textbf{v}\cdot \textbf{v}}{2}+\frac{\textbf{B}\cdot \textbf{B}}{8\pi},
\end{equation}
where the internal energy $\epsilon$ is given by 
\begin{equation}
\epsilon = \frac{p}{\gamma -1}
\end{equation}
and $\gamma = 5/3$. In our simulations, we will assume that the heat flux is confined to be parallel to the magnetic field lines. This 
assumption applies only when the ratio of electron gyro-radius to field gradient scales is small. Under this assumption, the  heat flux 
parallel to field lines can be written as
\begin{equation}
Q = -\kappa_{\parallel}(\nabla T)_{\parallel},
\end{equation}
where the subscript $||$ indicates parallel to the magnetic field, and $\kappa_{\parallel}$ is the classical Spitzer heat conductivity: 
$\kappa_{\parallel} = \kappa_c\,T^{2.5}$, with $\kappa_c=2\times10^{-18}\,cm\,s\,g^{-1}\,K^{-2.5}$. We take  $\kappa_{\parallel}$ to be 
a constant throughout our  simulations and so hereafter  write it simply as $\kappa$.

The ASTROBEAR code uses an operator splitting method to solve these MHD equations with heat conduction. The viscosity and resistivity are 
ignored in our calculation so the dissipation is numerical only.  The ideal MHD equations are then solved with the MUSCL (Monotone 
Upstream-centered Schemes for Conservation Laws) primitive method with TVD (Total Variation Diminishing) preserving Runge-Kutta temporal 
interpolation.  The result is then sent to the implicit linear solver utilizing the High Performance Preconditioners (Hypre) to solve the 
anisotropic heat conduction equation. The linear solver requires temporal sub-cycling technique to maintain its accuracy. The code runs in 
parallel with fixed grid domain.

\section{Problem Description and Analytical Model}
Our initial set up involves  hot  and  cold regions  separated by a thin planar interface. We  study how the magnetic field configuration 
alters the heat transfer rate between the hot and cold regions in presence of anisotropic heat conduction. We study the problem in  2-D. 
 
To  guide subsequent interpretation of the results, we first compare two simple but illustrative limits of magnetic field orientation: (1) a uniform 
magnetic field aligned with the direction normal to the interface; (2) a uniform magnetic field perpendicular to the normal direction of the interface. 
In case (1), because  the angle between the magnetic field and  temperature gradient is everywhere zero,  heat conduction across the interface 
is expected to take on the  Spitzer value associated with isotropic heat conduction. In case (2) however, the angle between the magnetic field and 
the temperature gradient is always $90^{\circ}$, so with our approximations,  heat cannot flow across the interface.

We define a heat transfer efficiency $\zeta$ equal to the magnetic field-regulated heat transfer rate divided by the isotropic Spitzer rate, namely, 
\begin{equation}
\zeta = \frac{q}{q_i}
\end{equation}
where $q$ is defined as the amount of thermal energy transported through the interface per unit time.   
The average angle $\theta$ between the temperature gradient and the uniform magnetic field then plays an important role in determining $\zeta$. 
At $\theta = 0$, $\zeta = 1$. At $\theta = \pi/2$, $\zeta = 0$.

We  now address  the influence of both a mean field and a tangled field $\zeta$. Consider there to be  a strongly tangled local field that has no mean 
value in the direction normal to the interface, i.e. ${\bf B}_{0,x}$ whose total magnitude is ${B}_0$, and  a global magnetic field ${\bf B}_d$ aligned 
with the normal of the interface of magnitude $B_d$. If $B_d \gg B_0$, the magnetic field around the interface only slightly deviates from the normal 
direction and $\zeta$ should be close to 1. If $B_d \ll B_0$, one would expect that the global energy transfer would be slow and  $\zeta$ should be 
close to zero.

If  $B_d$ and $B_0$ are comparable, we expect $0< \zeta <1$. We also expect $\zeta$ can change throughout the evolution if the strucuture of the 
magnetic field is  modified by the dynamics of heat transfer. It is instructive to ask whether the feedback from the magnetic field structure evolution will 
amplify the heat transfer by  creating more channels,  or shut it down. The answer depends on the influence of  magnetic reconnection, as we  will see 
from the simulations. Only if magnetic reconnection acts to smooth out local small scale structures and link the initially isolated structures to the global 
mean field across the interface then we would expect the heat conductivity to increase.

\begin{figure}
\begin{center}
\includegraphics[width=\columnwidth]{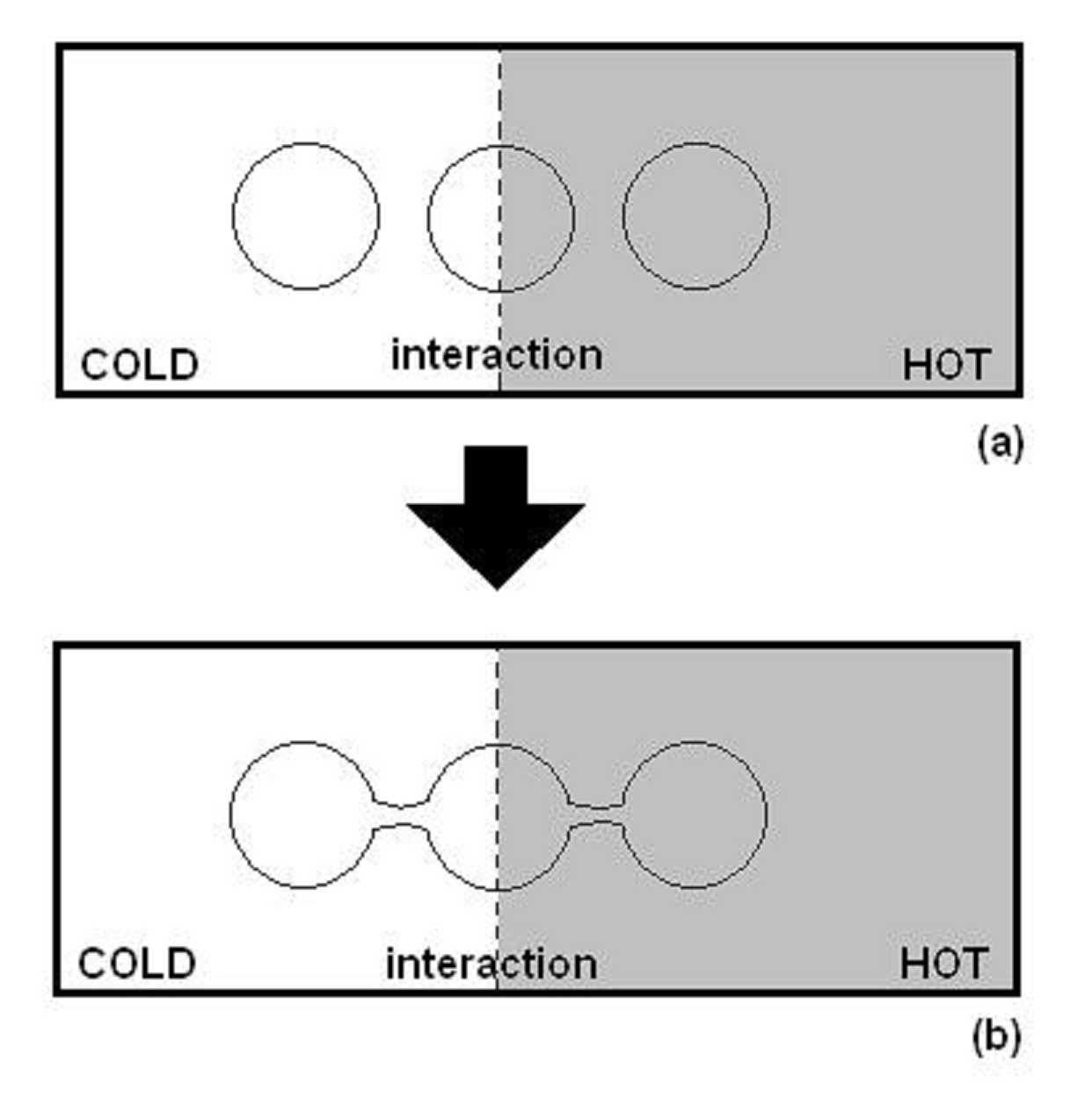}
\caption{The initial and steady state field configuration. (a): the initial field forms complete loops that only allows heat transfer within the interaction 
region. (b): the steady state field reconnects itself so that it allows heat transfer between regions deeply into the hot and cold areas. }
\label{fig01}
\end{center}
\end{figure}

\placefigure{fig01}

In what follows,  we refer to the initial tangled field region as "the interaction region". Figure~\ref{fig01}(a) shows a schematic of  initial and hypothetical evolved 
steady state field configurations for such a tangled field set up. From the figure we can see that the initial field configuration forms a ''wall'' which restricts 
energy transfer across the two  interaction region. However, if the subsequent evolution evolves to the  steady state shown in (b), then expansion of the 
interaction region and magnetic reconnection has allowed the field to penetrate through the entire region.  Thus the initial "wall" of tangle field  wall is destroyed 
and thermal conduction will be less inhibited than initially. We will check how accurately  this proposed picture of destruction of field wall is valid from analyzing  
our numerical simulations, and quantitatively discuss the effects on the energy transfer.

\section{Simulation Setup}
For our initial conditions, we  set up an interface between hot and cold regions  in mutual pressure equilibrium. The temperature distribution on the horizontal $(x)$ 
axis is given by:
\begin{equation}
T(x) = T_0(1-x^2)^{0.4}
\end{equation}
in the region  $0 < x < 1$ with $T_0 = 100$ in computational units. This temperature profile has a sharp gradient at $x=0$. The temperature distribution is 
plotted in Figure~\ref{fig02}(a). The region $0.4 < x < 0.5$ is the interaction region we described in the previous section. At the two side boundaries, the temperature 
is set to be constant, and uniform across the regions of each respective side of the box connecting to that side of the interaction region. We are  primarily 
interested in the region of the box where the heat transfer occurs and  noticeably evolves  during the simulation run time.  This means we  will mainly focus 
on the interaction region.  The horizontal length of the interaction region in the simulation domain is $0.1$ in computational units. 

The thermal pressure is set to be in equilibrium over the entire box, that is
\begin{equation}
P(x) = P_0
\end{equation}
with $P_0 = 100$. The density distribution is set up by the ideal gas law, namely:
\begin{equation}
\rho(x) = \frac{P(x)}{T(x)}
\end{equation}
in computational units.

For the Spitzer diffusion coefficient, we assume the diffusion is linear as in Eq.(7), and use the approximation: $\kappa_{\parallel} = \kappa_c\,T_{mid}^{2.5}$, 
where $\kappa_c$ is the classical conductivity, and $T_{mid}$ is taken to be the middle value of temperature across the interface, about $0.5\,T_0$.

We choose the initial field configuration:
\begin{equation}
B_x = B_d+B_0\,\sin(n\,\pi\,y/\lambda),
\end{equation}
\begin{equation}    
B_y = B_0\sin(n\,\pi\,x/\lambda)
\end{equation}
where $n$ and $\lambda$ are the mode number and wavelength of the tangled field respectively, $B_0 = 10^{-3}$ in computational units, and $B_d$ 
can assume various initial values that reflect the evolving global field as the result of reconnection. This initial field configuration is therefore one of  a 
locally tangled field surrounding the interface with one measure of the tangle given by:
\begin{equation}
R = B_d/B_0
\end{equation}
When $R = 0$, there are only locally confined field lines, whereas $R = \infty$ indicates a straight horizontal field without any "tangling". As  $R$ increases, 
the  relative fraction of  field energy corresponding to lines which  penetrate through the interaction region increases. In our simulations, we  consider 
cases with $R = 0.0,\,0.2,\,0.4,\,0.6,\,1,\,2,\,4,\,\infty$. Figure~\ref{fig02}(a), Figure~\ref{fig04}(a) and Figure~\ref{fig05}(a) show the magnetic field 
configuration for initial $R$ values of $0.0$, $0.4$, $1.0$. 

We note that our MHD approximation a priori implies that the electron gydroradius is much smaller than the length scale of one grid cell. Thus the dissipation 
scale and all field gradient scales are larger than the electron gyro-radius by construction in our simulations. 
 
We run simulations with typical resolution  of 2048 cells on the horizontal axis in fixed grid mode. Runs with doubled resolution showed  no significant differences 
compared to the standard resolution runs. We use  fixed boundary conditions at the $x$  boundaries: the pressure, density and temperature at the two ends 
are fixed to their initial values, as is the magnetic field.  We use  periodic boundary conditions for the y-axis boundaries.

There are five  parameters whose influence determine the simulation behavior  and guide interpretation of results:

\textbf{1. Plasma $\beta$.} $\beta\equiv {8\pi P\over B^2}$ has little effect on diffusion because even with very high values of the plasma $\beta$ used in the 
simulation, we are still in the MHD regime and the gyro-radii of electrons are assumed small.  Thus the direction of thermal conduction is not locally affected 
by $\beta$.  It is possible that instabilities could arise in the low $\beta$ limit that affect pressure balance during the evolution of the simulations but that 
turns out not to be the case for the $\beta$ range of $10^{5} \sim 10^{8}$ that we use. The value of $\beta$ in this range does not exihbit any influence 
on the simulation result as indicated by our numerical experiment.

\textbf{2. Initial Tangle measure $R = B_d/B_0$.} If $R>>1$, the local small scale field can mostly be ignored and Spitzer  thermal conductivity is expected, 
whereas if $R<<1$ a value much less than Spitzer is expected.

\textbf{3. Ratio of the diffusion time scale to the sound crossing time scale for one grid cell:}

\begin{equation}
r =  t_{diff}/t_{hy}=\frac{\rho\,C_s\,l}{\kappa_{\parallel}}
\end{equation}
where $\rho$ is the density, $l$ is the characteristic gradient  length scale of temperature: $l = min(\frac{T}{|\nabla\,T|})$ and $C_s$ is the sound speed. If 
$r<<1$, thermal diffusion would initially  dominate  and the pressure equilibrium would be broken by this fast energy transfer. If $r>>1$, then the pressure 
equilibrium would be well maintained throughout the entire evolution and the energy transfer may be viewed as a slow relaxation process. In our simulation, 
$r \approx 0.3$ initially, so that  diffusion induces a pressure imbalance. Eventually,  as the heat transport slows, the pressure equilibrium catches up and is 
maintained.

\textbf{4. Ratio between the temperature gradient scale length and the wavelength of the tangled field: $h = 2\,\pi\,l/\lambda = k\,l$} . If  $h = 0$ there is 
no tangled field, and no inhibition to  heat transfer. As $h$ inreases, the field becomes more tangled, and the energy is harder to transfer. However, a large 
$h$ value may also result in increased  magnetic reconnection, because the Lundquist number of field confined in a smaller region is larger, for the same field 
strength. Thus would then lower $h$.

\textbf{5. Mean global energy transfer rate: $q = \delta E/t_{bal}$}, where $t_{bal}$ is defined as the time needed for the hot region and cold region to reach 
a certain degree of temperature equilibrium  by a transfer of heat energy $\delta E$ across the interface.

A mathematical expression for the heat transfer rate can be derived by considering a slab with  a planar  interface   aligned with the $y$ direction at the middle 
of the  interaction region  $(x=0)$ with a tangled magnetic field, and an average temperature gradient aligned in  the $x$ direction.  Define the global temperature 
gradient as $|\nabla\,T|_{g} = (T_{hot}-T_{cold})/(T_0\,L)$, where the subscripts ``hot'' and ``cold'' denote the characteristic temperatures of the hot and 
cold regions, $L$ is the width of the interaction region, and  $T_0$ is a normalization factor which ensures $|\nabla\,T|_{g}$ has dimensions of inverse length. 
We can then integrate over the  volume of the interaction region (and since  there is no $z$-dependence, the essential content is an area integral)  to obtain the 
effective heat flux through this region:
\begin{equation}
    \overline{q} = D\,|\nabla\,T|_{g}\int{\frac{B_d}{|B|}\,dx\,dy},
\end{equation}
where $D$ is a constant that depends on neither the magnetic field nor the temperature distribution, $|B|$ is the local field strength. Notice that this expression is 
valid only when the magnetic field is varying at a length scale smaller than the interaction region length.

Using Eqs.(12), (13) and (14) in (16) and the approximation that  the areal average in the interaction region 
 $\langle {\bf B}_0 \cdot {\bf B}_d\rangle \sim 0$ so that 
 $\langle ( {\bf B}_0 + {\bf B}_d)^2 \rangle \sim \langle  B_0^2+ B_d^2 \rangle$,
we obtain
\begin{equation}
\overline{q} \approx D\,\frac{|\nabla\,T|_{g}\,R}{\sqrt{1+R^2}}.
\end{equation}
For the unmagnetized isotropic  case, or for transfer with a field entirely aligned with the temperature gradient,  we have instead
\begin{equation}
\overline{q}_i = D\,|\nabla\,T|_{g}.
\end{equation}

Dividing Eq.(18) by Eq.(19), we obtain an appoximation for the heat transfer efficiency over the interaction region:
\begin{equation}
\zeta = \frac{R}{\sqrt{1+R^2}}.
\end{equation}
If the initial temperature profiles are identical for different field configurations, this formula can then be used to estimate the expected energy tranfer rate from 
situations with various field configuration. By normalizing the heat transfer rate to that of the isotropic heat conduction case, we  obtain the heat transfer efficiency 
$\zeta$. The accuracy of Eq.(19) can be tested by plotting the heat transfer efficiency obtained from the simulations against measured values of $R$.

\paragraph{}If magnetic reconnection occurs during the time evolution of the heat transfer process, then conduction channels can open up and the energy 
exchange  can be accelerated. We would then expect the actual curve of $\zeta$ vs $R$ to evolve to be higher than the value Eq.(19) predicts in situations with low 
$R$ values. Meanwhile, for high $R$, the analytical prediction and the real physical outcome should both approach the horizontal line $\zeta = 1$, which denotes 
conductive efficiency consistent with the unmagnetized case.  We emphasize that $R$ as used in this paper is  always calculated with the the initial values of the 
magnetic field, not time evolved values, and that Eq.(19) is valid when estimating a cold to hot interface with initial tangle measure as the ratio of initial global straight 
field to initially local tangled field. To follow a measure of the tangle that evolves with time, a generalized  tangle measure should be calculated in a more sophisticated 
manner and the integral form (Eq.(16)) should be applied.

\section{Simulation Results}

\begin{figure}
\begin{center}
\includegraphics[width=\columnwidth]{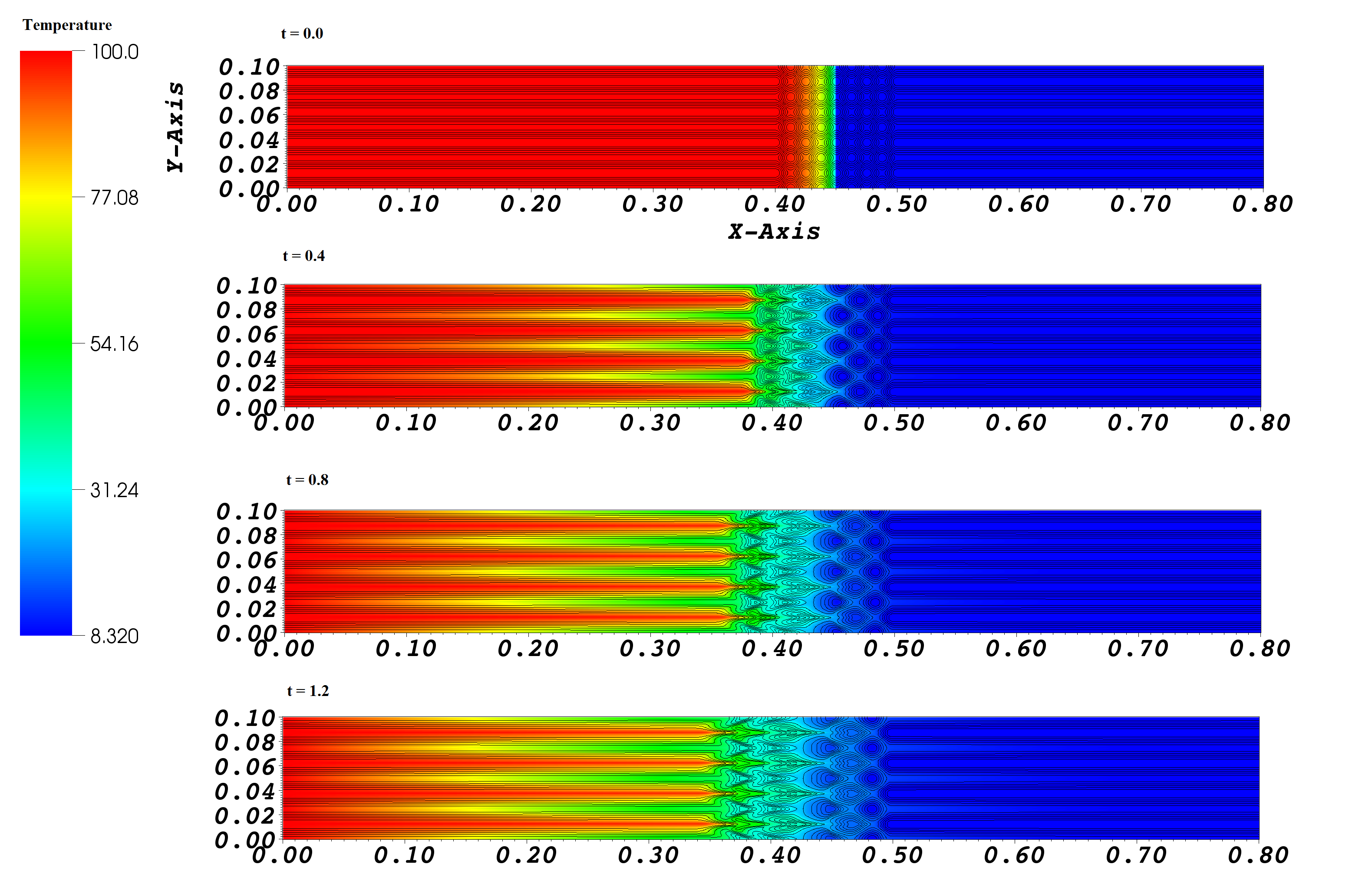}
\caption{Evolution of temperature distribution with $R = 0.0$. The cuts are at (a): $t=0.0$, the initial state, (b): $t=0.4$, (c): $t=0.8$, (d): $t=1.2$, 
the steady state.}
\label{fig02}
\end{center}
\end{figure}

\begin{figure}
\begin{center}
\includegraphics[width=\columnwidth]{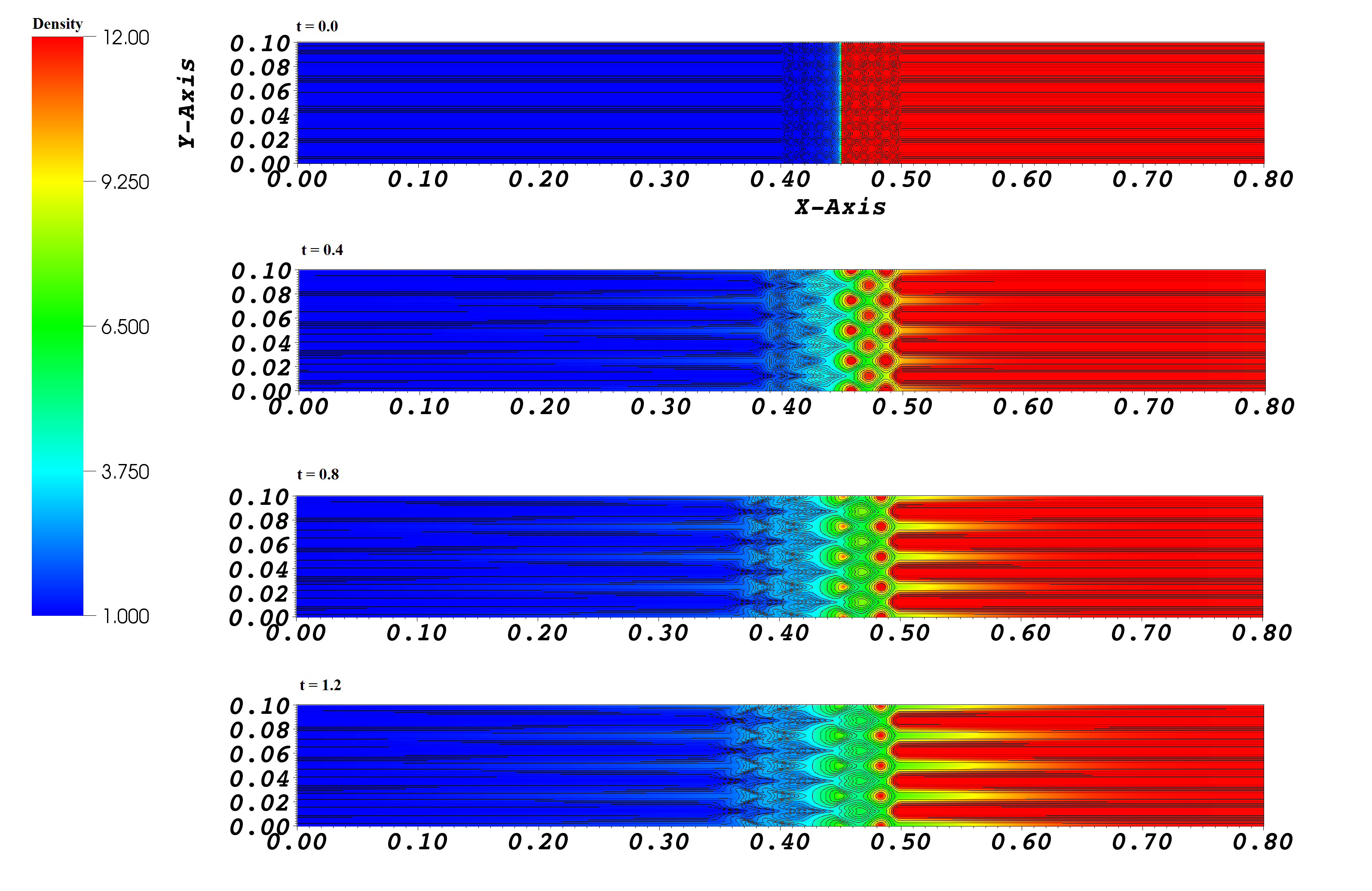}
\caption{Evolution of density distribution with $R = 0.0$. The cuts are at (a): $t=0.0$, the initial state, (b): $t=0.4$, (c): $t=0.8$, (d): $t=1.2$, the steady state.}
\label{fig03}
\end{center}
\end{figure}

\begin{figure}
\begin{center}
\includegraphics[width=\columnwidth]{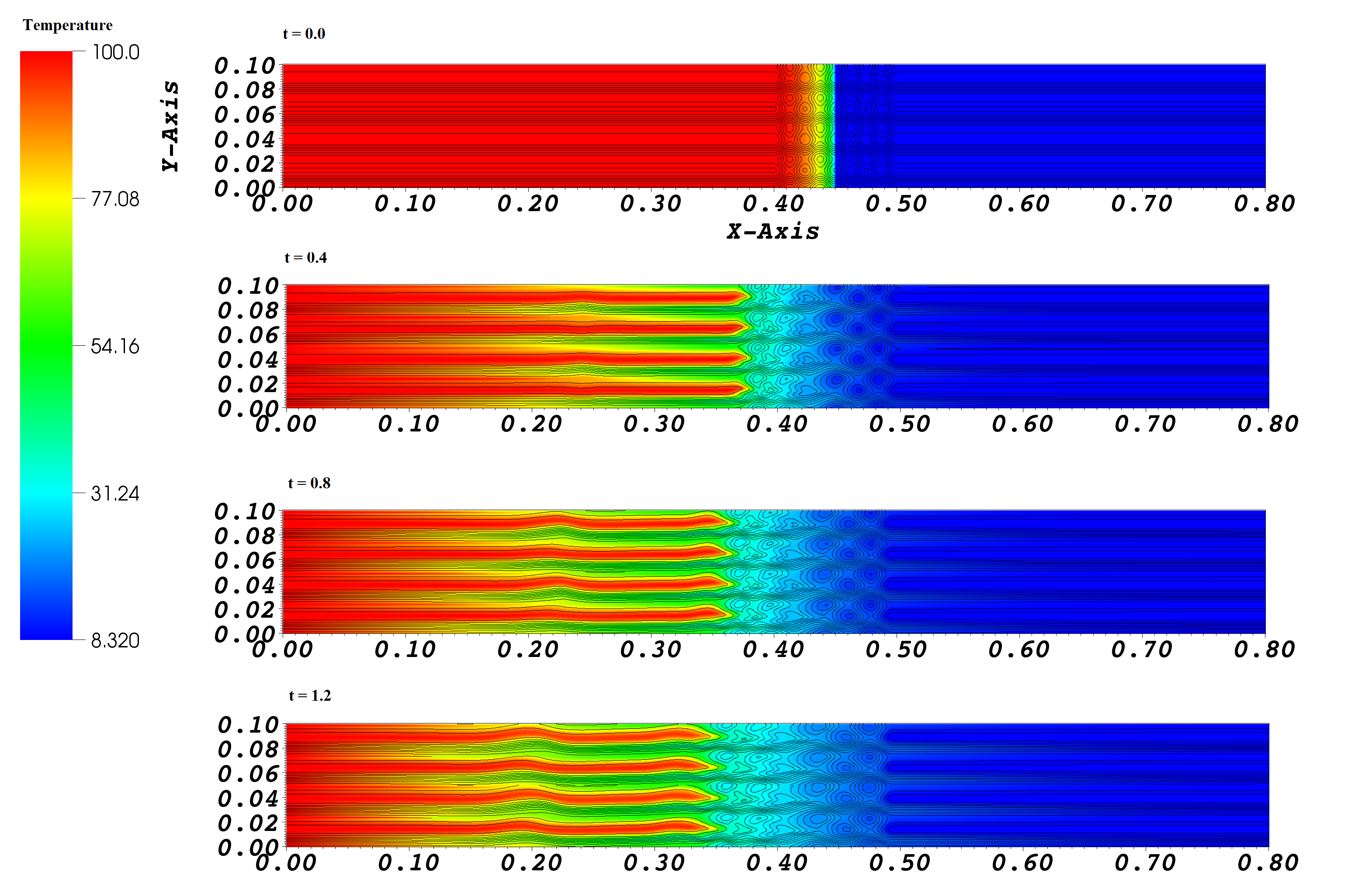}
\caption{Evolution of temperature distribution with $R = 0.4$. The cuts are at (a): $t=0.0$, the initial state, (b): $t=0.4$, (c): $t=0.8$, (d): $t=1.2$, the steady state.}
\label{fig04}
\end{center}
\end{figure}

\begin{figure}
\begin{center}
\includegraphics[width=\columnwidth]{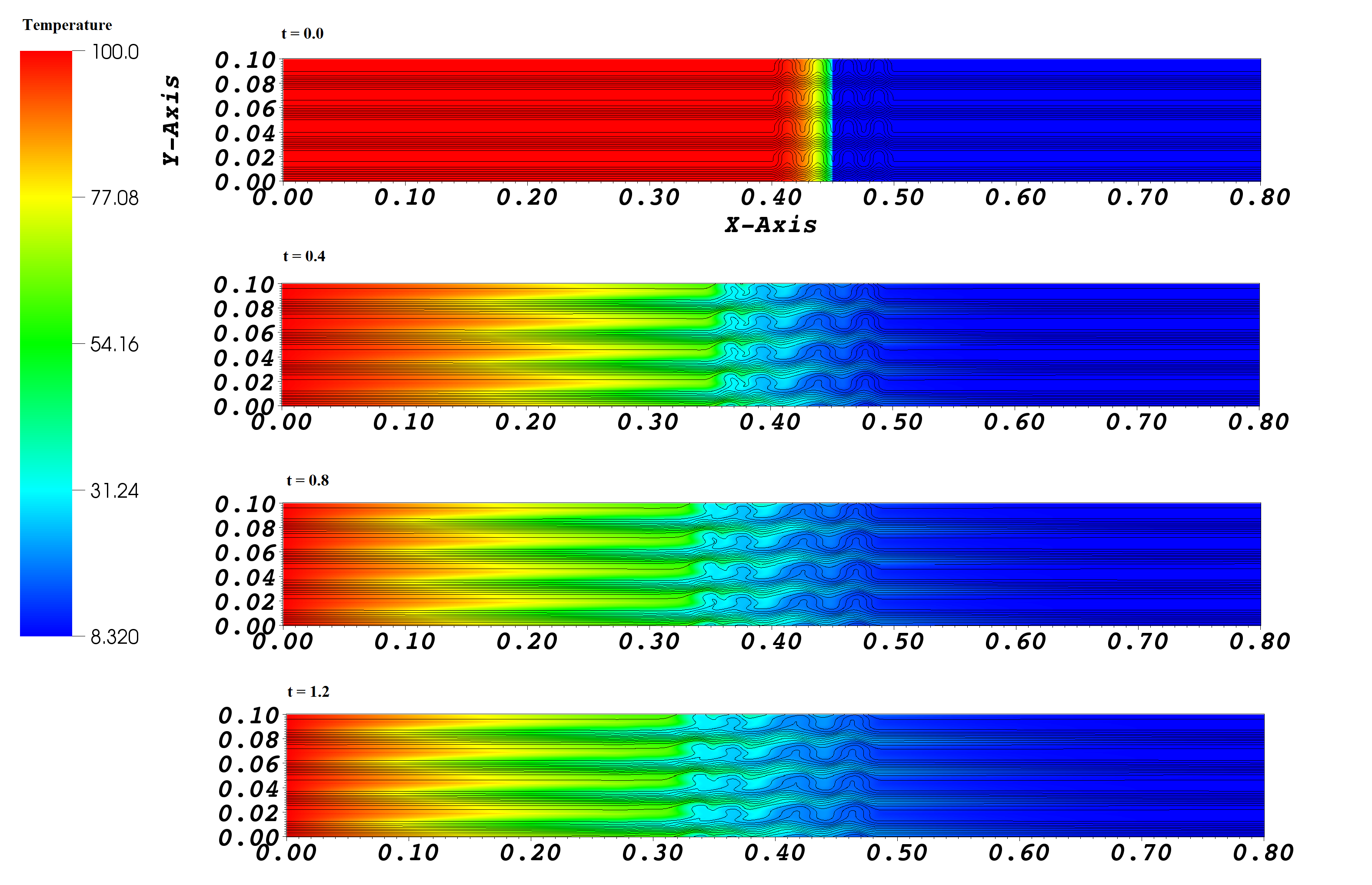}
\caption{Evolution of temperature distribution with $R = 1.0$. The cuts are at (a): $t=0.0$, the initial state, (b): $t=0.4$, (c): $t=0.8$, (d): $t=1.2$, the steady state.}
\label{fig05}
\end{center}
\end{figure}

We choose initial conditions with values $R = 0.0,\,0.2,\,0.4,\,0.6,\,1.0,\,2.0,\,4.0$ to run the simulations. The simulation run time is taken to be $1.2$ (which 
corresponds to $12,000$ years in real units for WBB. The initial cuts of temperature and magnetic field lines for $R = 0.0,\,0.4,\,1.0$ are shown in Figure~\ref{fig02}(a), 
Figure~\ref{fig04}(a) and Figure~\ref{fig05}(a) respectively. Figure~\ref{fig03}(a) shows the initial cut of the density distribution in the $R=0.0$ run. 
We also run simulations with purely horizontal 
magnetic field lines, equivalent to the $R =\infty$ case, and runs with purely vertical field lines. Frames (b) to (d) in Figure~\ref{fig02} to Figure~\ref{fig05} are 
from the late stages of the evolution, and the final frames always display the steady state of the runs.  A steady state is facilitated by the fact that the  boundaries 
are kept at a fixed temperature throughout the simulations.

\placefigure{fig02}

\begin{figure}
\begin{center}
\includegraphics[width=.9\columnwidth]{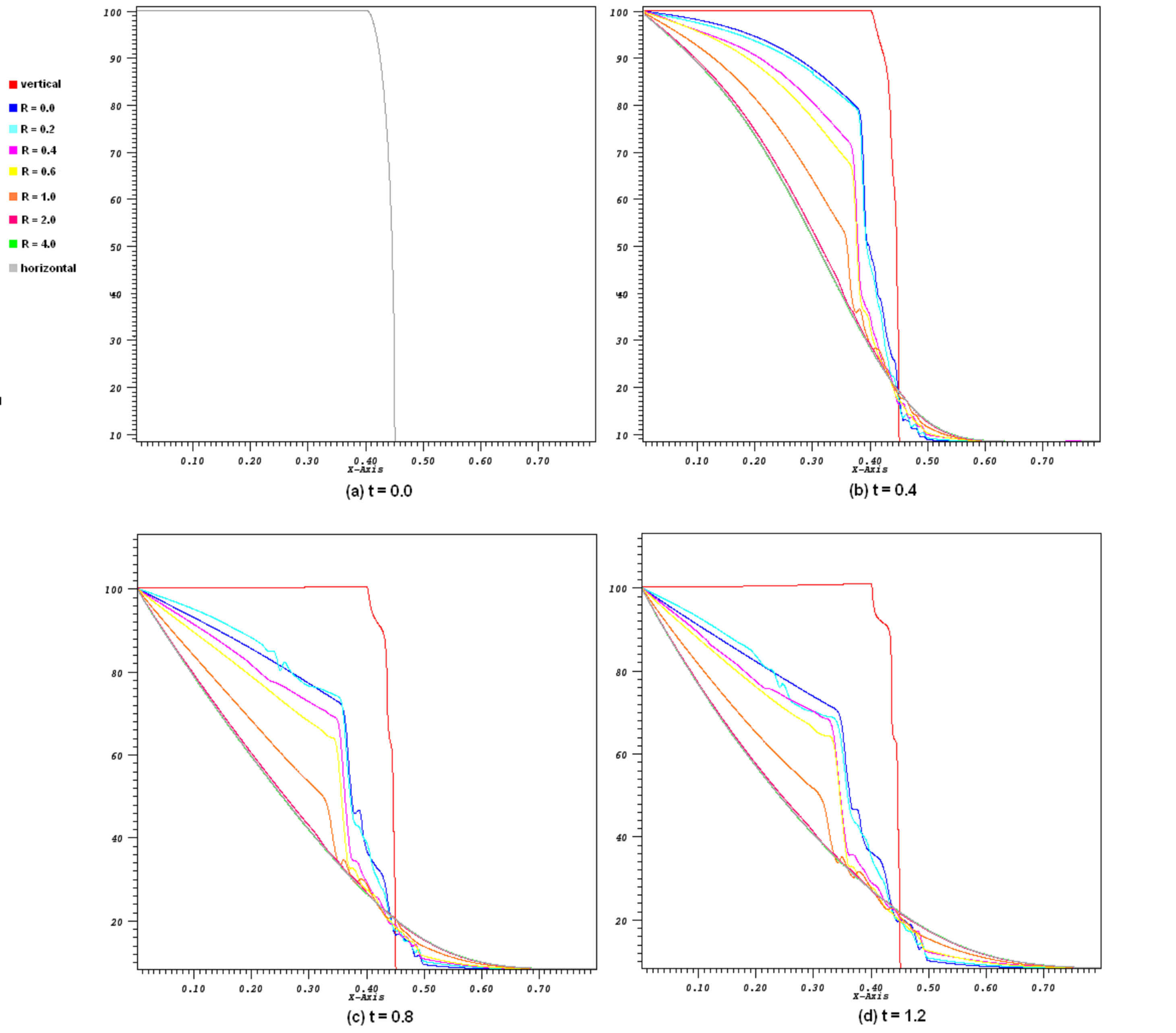}
\caption{Evolution of mean cut temperature averaged on y direction with different $R$ values labeled by different colors. The cuts are at (a): $t=0.0$, the initial state, (b): $t=0.4$, (c): $t=0.8$, (d): $t=1.2$, the steady state.}
\label{fig06}
\end{center}
\end{figure}

In Figure~\ref{fig06}, we plot the mean cuts of the temperature $T_c$,   obtained by averaging the temperature along $y$ axis, against the $x$ position for selected 
evolution times. Since  the anisotropic heat conduction is initially faster than the 
pressure equilibration rate, the energy distribution around the temperature interface change rapidly until about $t = 0.4$. This energy transfer is mostly 
confined to the interaction region for the low $R_0$ runs, since in these cases only a few field lines can penetrate into the entire interaction region. 

\placefigure{fig03}

\placefigure{fig04}

\placefigure{fig05}

During the initial heat exchange phase, the thermal energy and density quickly redistribute in the interaction region. As seen in Figure~\ref{fig02}(b), islands at 
$x=0.48$ are formed by material bounded by the magnetic field lines, since the field orientation blocks heat exchange with the surroundings. Around 
$x=0.4$, there are also cavities formed where the thermal energy is inhibitted  from flowing. The magnetic field lines, which form complete sets of loops 
in the $R=0.0$ case, begin to distort. It can be observed that the field lines are more strongly distorted  in the low density part of the interaction region 
than in the high density part.  This occurs because velocity gradients are driven by the early rapid redistribution of heat (pressure) by conduction.

At time $t=0.4$ (see Figure~\ref{fig02}(b)), the field lines surrounding the cavities at $x=0.4$ reconnect, making thermal exchange possible. During the evolution, field 
lines begin to link the interaction region to the hot material on the left. This phenomenon is most apparent in Figure~\ref{fig02}(d), which marks the final state of the 
thermal energy exchange. We also see that there is little difference between Figure~\ref{fig02}(c) and Figure~\ref{fig02}(d), because at late stage of the process, the thermal 
diffusion gradually slows  so that the magnetic field configuration approaches a steady state. 
 
By comparing Figure~\ref{fig06}(c) with Figure~\ref{fig06}(d), we see that the  mean cuts of temperature show little difference for all  values of $R$. The mean cuts of temperature 
$T_c$ exhibit a jump in the region of $x=0.35\,\sim\,0.5$, but are relatively smooth on either side of this region. This shows that even though the 
tangled field "wall" has been broken and allows channels of thermal conduction through it,  the temperature profiles is not as smooth as  in the purely 
straight field case.

\placefigure{fig06}

For the cases of $R=0.4$, there are field lines which penetrate the entire interaction region from the start. By observing the evolution of the magnetic 
field lines at about $x=0.38$, we see that magnetic reconnection is still happening, and causes the field loops to merge. The  observed behavior 
resembles the process displayed by Figure~\ref{fig01}. When $R=1$, there are hardly any  temperature islands that  bounded by magnetic field loops. 
The evolution of the field lines shows less dramatic reconnection and evolve in what appears as more gentle straightening.

\section{Discussion}
\begin{figure}
\begin{center}
\includegraphics[width=.9\columnwidth]{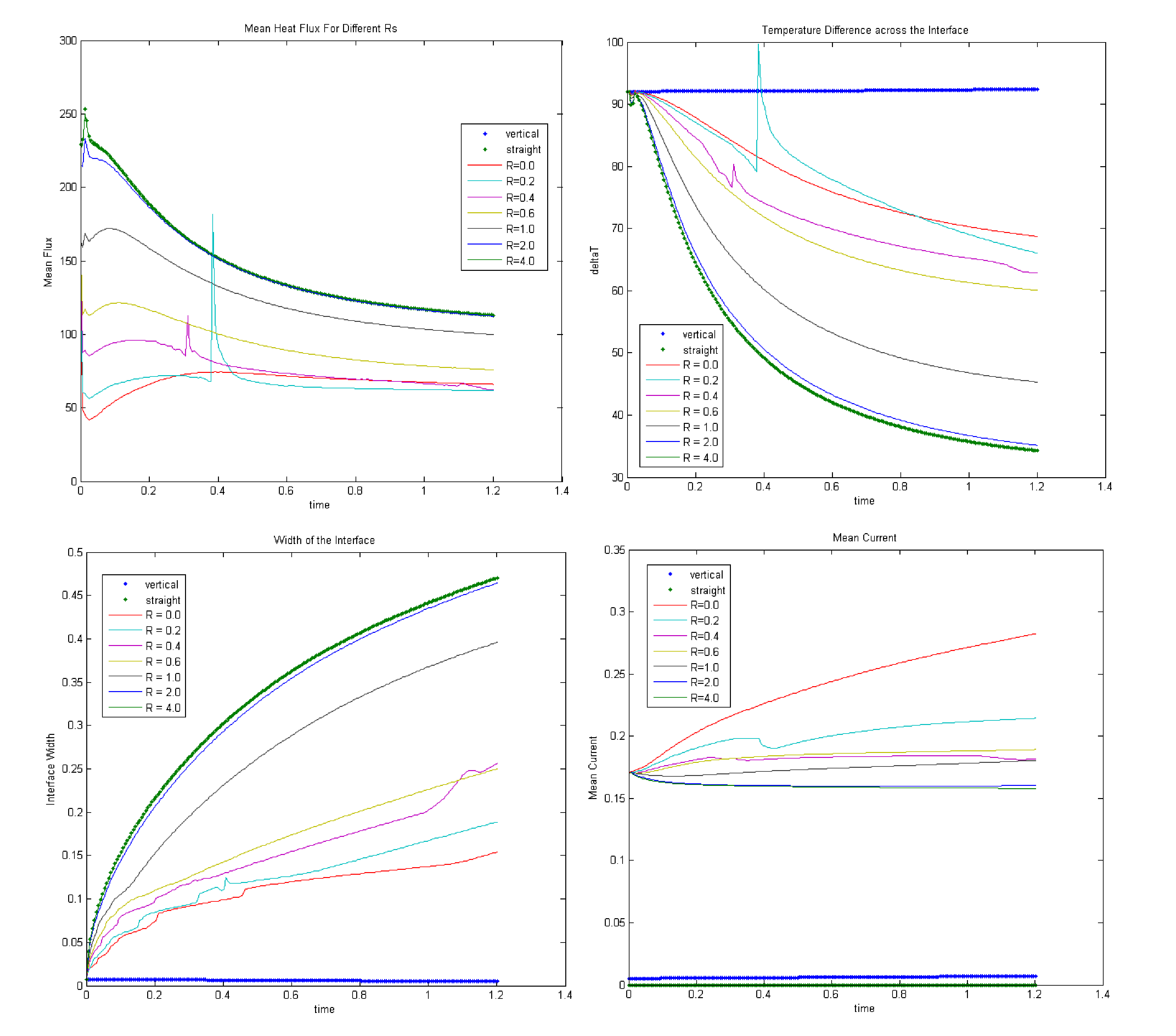}
\caption{(a) top left: time evolution of mean heat flux at the interface, (b) top right: time evolution of average temperature difference between the hot and cold regions, (c) bottom left: time evolution of interface width, (d) bottom right: time evolution of the mean value of $|\curl\textbf{B}|$.}
\label{fig07}
\end{center}
\end{figure}

\placefigure{fig07}
We begin our analysis with the evolution of the heat flux. The average heat flux per computation cell for different values of $R$ is plotted as a 
function of time in Figure~\ref{fig07}(a). Note that in the vertical field case (${\bf B} = B_y {\hat y}$) the heat flux remains zero as field entirely inhibits electron 
motion across the interface. For cases with $R > 1$, the heat flux decreases throughout the evolution. Recall that $R > 1$ implies cases where the 
"tangled" portion of the field is relatively weak and heat is quickly transported from one side of the interface to the other.  Thus the trend we see 
for $R>1$ occurs as the temperature distribution approaches its equilibrium value. For lower $R$ values, especially those of $R < 0.5$, an initial 
phase of heat flux amplification is observed as magnetic reconnection in the early evolution opens up channels for heat to transfer from hot to cold 
regions. At the late stage of the evolution when reconnection has established pathways from deeper within the hot region to deeper within the cold 
region temperature equilibration dominates leading to a decreasing heat flux phase as observed in the $R > 1$ cases. Note that the similarity between 
the $R > 2$ cases and the $R = \infty$ case is predicted by Eq.(19): as the global field comes to dominate, the heat flux inhibition imposed by anisotropic 
heat conduction in the local tangled field  can be ignored.

\begin{figure}
\begin{center}
\includegraphics[width=.6\columnwidth]{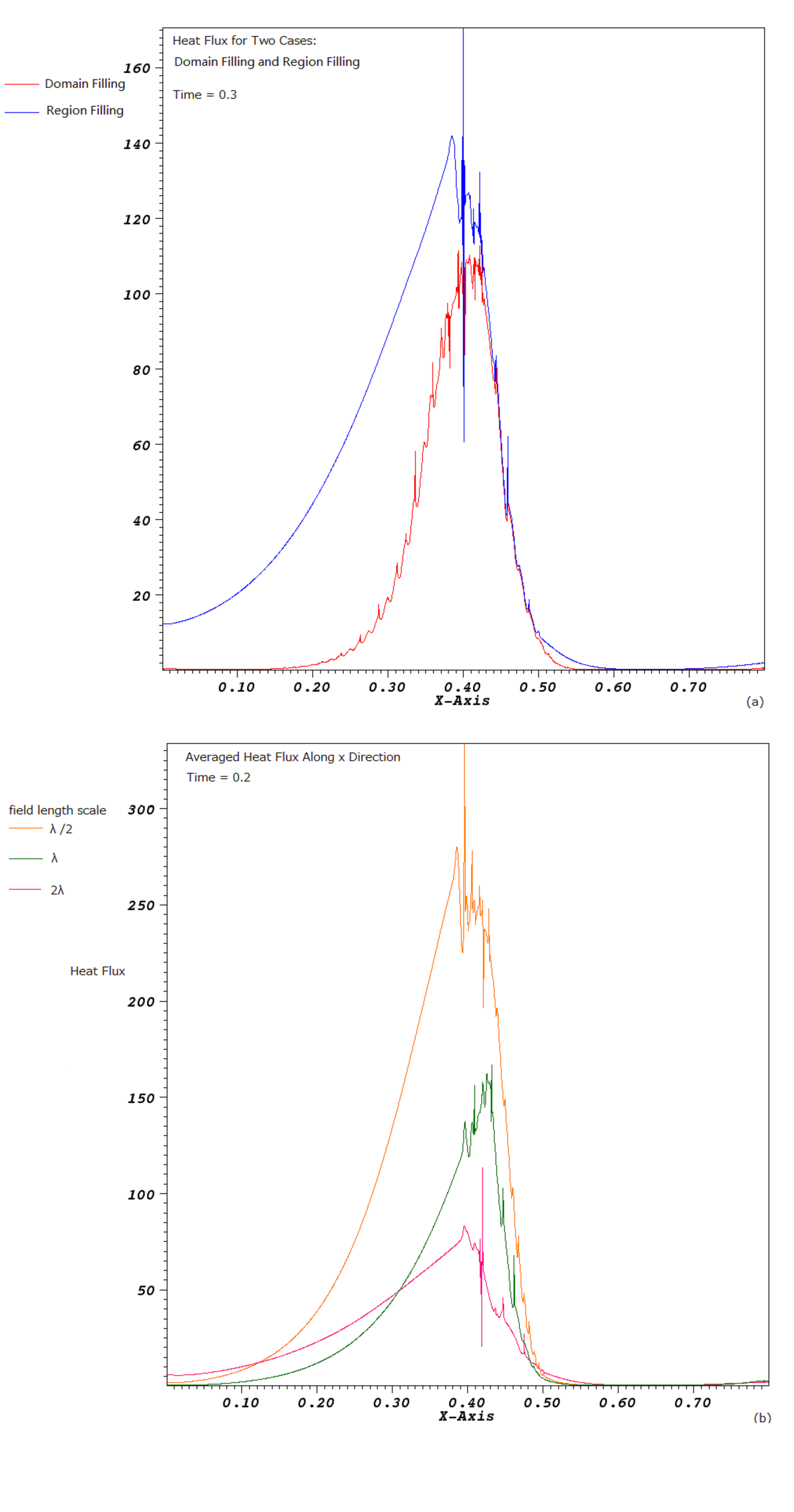}
\caption{(a) Comparison of averaged heat flux for situation with field loops filling up the entire domain and situation with field loops only fill the interaction region. (b) Comparison of averaged heat flux for situations with different tangeld field length scale.}
\label{fig08}
\end{center}
\end{figure}

\placefigure{fig08}

In order to understand the influence of magnetic reconnection on heat transfer rates we compare simulations with different filling fractions of the 
tangled field.  Two cases are shown in Figure~\ref{fig08}(a): (1) a temperature interface with a "volume filling" tangled field and (2) temperature interface with 
the tangled field filling only the region surrounding the interface. In case (2) the rest of the domain is filled with straight field lines connecting the hot 
and cold regions.  From Figure~\ref{fig08}(a) we see that case (1) shows much slower heat transfer rates compared to what is seen in case (2).  This results 
because reconnected field lines in case (2) are linked to the globally imposed background field that in turn linking the hot and cold reservoirs.  In case (1) 
reconnection only  leads to larger field loops but cannot provide pathways between the reservoirs. The effect of different scale lengths on the  evolving 
field loops is shown in Figure~\ref{fig08}(b) in which we plot the result from three simulations wavelengths for the tangled field component (tangled field "loops").  
Note that  $\lambda$ is defined in Eq (12) and Eq.(13). We use a sequence of values for wavelength: $2\lambda$, $\lambda$ and $\lambda/2$. 
Figure~\ref{fig08}(b) clearly shows that smaller field loop $\lambda$ leads to the largest average heat flux, since smaller scale loops will reconnect before large loops 
for a given magnetic resistivity. This result demonstrates the link between the number of reconnection sites  of the field and heat flux.

We next analyze the temperature equilibration in detail. The averaged temperature difference across the interface is plotted in Figure~\ref{fig07}(b). It shows the 
difference between the averaged temperature at the hot side and the cold side. One significant feature in Figure~\ref{fig07}(b) is that the temperature difference 
decreases to a steady value $T_{end}$ in all cases. This resembles percolation across  a membrane which allows a density jump to happen when filtering 
two fluids. Figure~\ref{fig07}(c) shows the distance required for the temperature to drop $80$ percent at the interface. This distance characterizes the length of the 
interaction region. Except for the vertical field case where no heat transfer is allowed, the interface is expanding at different rates for different $R$ values. 
The expansion for all the cases of nonzero $R$ approaches a steady value which is also a characteristic feature of the temperature equilibration evolution.

We now analyze the modification of magnetic field configuration during the evolution. Throughout our simulations, the local magnetic field is initially a 
set of complete loops surrounding the interaction region. Once the energy transfer begins, the interaction region tends to expand as discussed previously. 
This expansion stretches the field lines on the x direction and  distorts these circular loops, eventually inducing  magnetic reconnection which oppens up 
channels connecting the hot and cold regions. From the current $J_B = |\curl \textbf{B}|$, we can get information on how tangled the field is. Figure~\ref{fig07}(d) 
shows the evolution of the mean value of the strength of $\curl \textbf{B}$ in the interaction region. We observe that in the vertical and straight field case, 
$|\curl \textbf{B}|$ remains constant, but decreases to a fixed value for $R \geq 2$ cases. This means the field in high $R$ cases is straightened by the 
stretching of the interaction region as seen in Figure~\ref{fig07}(c). For the $R \leq 1$ cases, we see that  $|\curl \textbf{B}|$ increases.  This rise is due to magnetic 
energy brought in via the cold mass flow and the creation of fine field structures that amplify $J_B$ faster than dissipation caused by interface expansion.

\begin{figure}
\begin{center}
\includegraphics[width=.7\columnwidth]{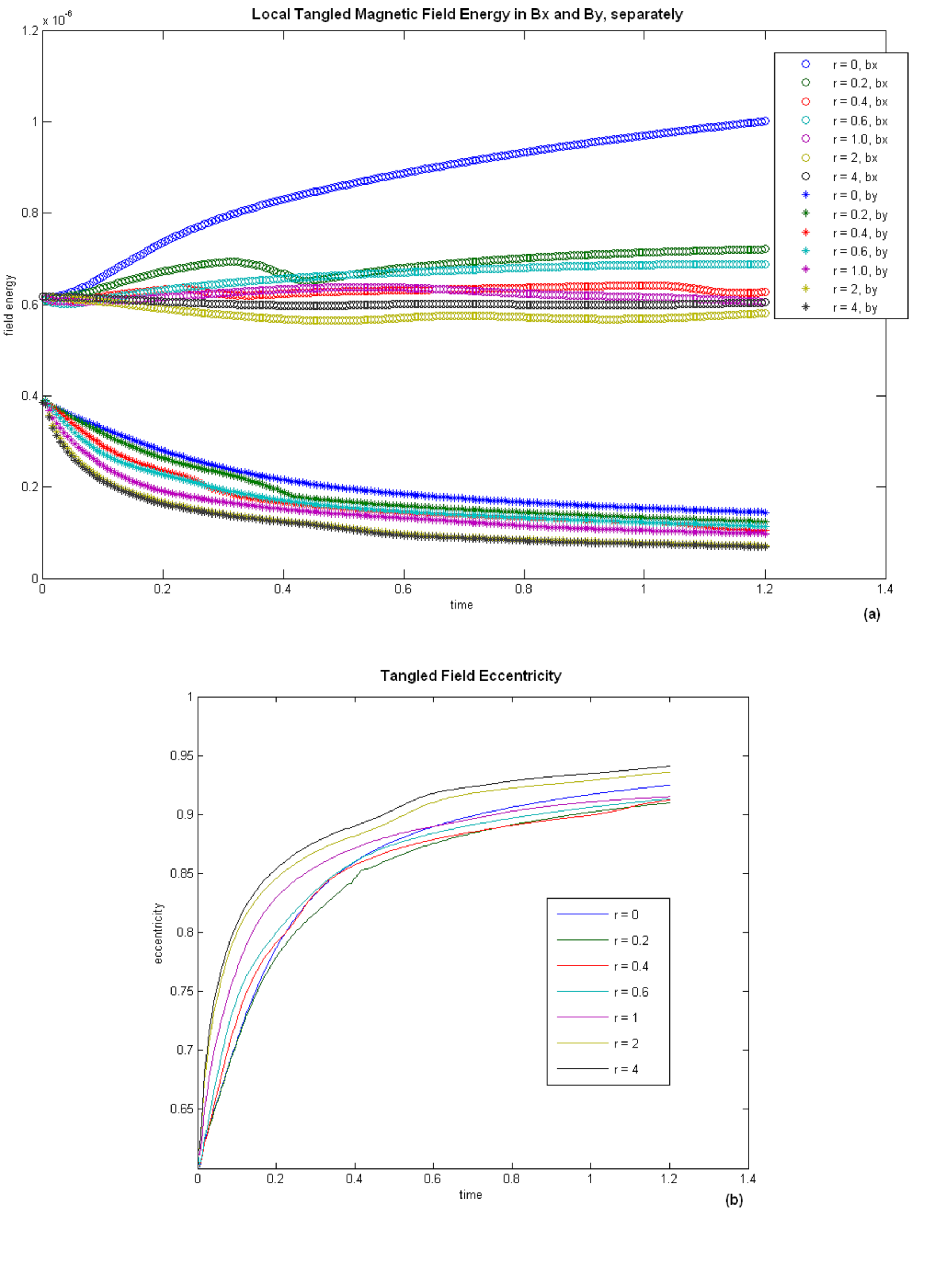}
\caption{(a) Comparison on evolution of local field energy in terms of $B_x$ and $B_y$. Circles corresponds to the $B_x^2/2$ curve, stars corresponds to the $B_y^2/2$ curve. The different colors denote various $R$ values. (b) Eccentricity of the ellipses constructed by assigning the mean values of local $|B_x|$ and $|B_y|$ to the major and minor axes, respectively. The set of curves show different evolution patterns for different $R$ values.}
\label{fig09}
\end{center}
\end{figure}

\placefigure{fig09}

The local field distortion can be clearly demonstrated by studying the energy evolution of magnetic energy stored in different field components. 
In Figure~\ref{fig09}(a), we plot the evolution of mean magnetic energy stored in the vertical field $\bar{B}_y^2/2$, compared with $\bar{B}_x^2/2$.  We note 
that the latter includes only the fluctuating contribution to the energy in the x field-- that is, the contribution to the horizontal field that does not 
come from the global mean $x$ component.
 
From Figure~\ref{fig09}(a), we observe that the $B_y^2$ energy decreases while the $B_x^2$ energy either increases or remains the same for all cases. 
The magnetic energy evolution can thus be viewed as a conversion of vertical field to horizontal field. This conversion need not conserve the total 
magnetic energy of the local tangled field because of magnetic reconnection and because  material advecting magnetic field can flow in and out of 
the interaction region. By comparison, in the $R > 1$ cases, the thermal energy and local magnetic energy can both decrease and add to the kinetic 
energy of the material surrounding the interface, because of the fast thermal diffusion enabled by the strong global field.

The distortion of the local field loops can also be demonstrated by plotting the mean eccentricity of the field loops. In Figure~\ref{fig09}(b), we plot the mean 
eccentricity evolution. For all cases, the mean eccentricity is zero initially because of the circular shape of the field loops. Later in the evolution, 
large $R$ cases tend to evolve into a state of large eccentricity in the steady state. This is caused by  a rapid  expansion of the interface induced 
by the strong global field. In short,   large $R$ induces more distorted local field loops and less tangled total field due to fast interface expansion, 
while small $R$ values results in less eccentric local field loops but with more tangled total field and strong magnetic reconnection.

\begin{figure}
\begin{center}
\includegraphics[width=.8\columnwidth]{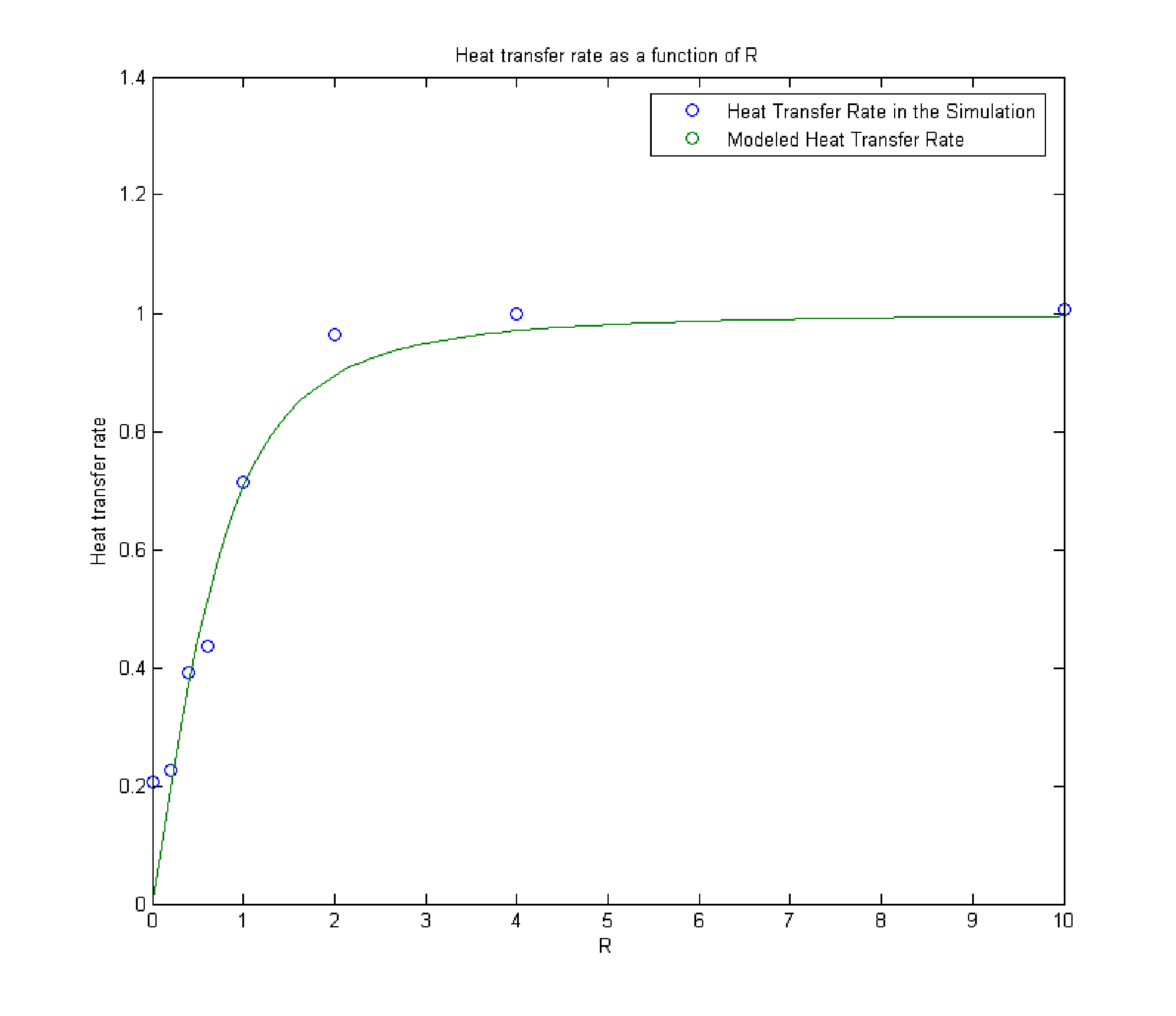}
\caption{heat transfer rate observed in the simulation compared with the analytic model.}
\label{fig10}
\end{center}
\end{figure}

\placefigure{fig10}

To compute the estimated heat transfer rate in the simulation, we calculate the averaged slope of the curve plotted in Figure~\ref{fig07}(b), and compare it to the 
analytic model in Section 4. Although the equilibration rate represented by the slope of the curves in Figure~\ref{fig07}(b) is changing throughout the evolution, 
an early phase of the evolution can be chosen when the field configuration has not been modified significantly for which we can then comptute the 
averaged heat transfer rate. By normalizing the resulting heat transfer rate to the isotropic value, we can determine the heat transfer efficiency 
for different magnetic structures. From Figure~\ref{fig10}, we can see that the analytic prediction and the simulation results agree quite well except for the situation 
when $R$ is below $0.2$. The simulation result does not converge to point (0,0) but ends at an intercept  on the $y$ axis. This intercept indicates that 
even if there are initially negligibly few channels for energy transfer, the magnetic reconnection can open up channels and allow heat transfer. Eq.(19) 
is valid for predicting the cooling rate of the hot material throughout the early phase of the heat equilibration process. It also provides insight on the 
strength of the local field in the vicinity of  the interface once we know the cooling rate and global magnetic field strength.

To summarize our results we find first that the average heat flux at the end of our simulations is lower than at the beginning for  all $R$ values.  Thus we see 
an approach to thermal equilibrum. In some cases we also see that the heat shows an initially increasing phase denoting a period of active magnetic reconnection. 

In the simulations we see the average temperature difference decreases to a constant value $T_{end}$ which is related to $R$. We also see the width 
of the initial interface expand to a fixed value during the simulation. 

Analysis of the simulation behavior shows that $J_B$ is an accurate measure of structural change in the magnetic field. Current decreases to a constant 
value for large $R$ cases and increases to a constant value for small $R$ values. 

Finally we have shown that Eq.(19) can be used to estimate the energy transfer rate for an initially complicated field structure by considering the relative 
strength of the local field and the global field. For those cases for which $R$ approaches $0$, Eq.(19) becomes invalid since the energy transfer in is mainly 
induced by a feedback from the magnetic field reconnection. By comparing cases with different field loop length scales, we demonstrate that the smaller 
the field loop length scale, the faster the reconnection rate. 

\section{Astrophysical Applications}

\placetable{tab01}

The issue of magnetized conduction fronts and their mediation of temperature distributions occurs in many astrophysical contexts. One long-standing 
problem that may involve anisotropic heat conduction are hot bubble temperatures in Wind Blown Bubbles (WBB).  WBB's occur in a number of setting 
including the Planetary Nebula (PN), Luminious Blue Variables (LBVs) and environments of Wolf-Rayet stars.  When a central source drives a fast wind 
($V_{wind} \sim 500 km/s$) temperatures in the shocked wind material are expected to be of order $10^7$ K, which is  greater than $2\,kev$.  The 
temperatures observed in many WBB hot bubbles via from X ray emission are, however in the range of  $0.5\,kev$ to $1\,kev$ range.  NGC 6888 is a 
particularly well known and well studied example for a WR star (\citet{zhe10}).  For planetary nebulae, Chandra X-ray observations have found 
a number of WBB hot bubbles with temperatures lower than expected based on fast wind speeds (\citet{mon05}, \citet{kas08}).  The role of 
wind properties and heat conduction in reducing hot bubble temperatures has been discussed by a number of authors (\citet{ste08}, \citet{aka07}, 
\citet{stu07}).  The role of magnetic fields and heat conduction was discussed in \citet{sok94}.

While our simulations herein were meant to be idealized  experiments aimed at identifying basic principles of anisotropic heat conduction fronts, we can 
apply  physical scales to the simulations in order to make contact with WBB evolution.  Tab.1 shows the results of such scaling.    Upon doing so, we 
infer that: $(1)$ given field strengths expected for WBB's, heat conduction is likely to be strong enough to influence on the temperature of the expanding 
hot bubble and the cold shell bounding it.  We also note that magnetic fields in WBB (for PN field strengths  see \citet{wou06}) are usually in the milli-Gauss 
range, and are relatively much stronger than the field strength that can be scaled to our simulations. Thus the magnetic field in realistic WBBs is highly likely 
to result in anisotropicity and regulate the behavior of heat conduction. Since the heat transfer does not directly depend on the magnetic $\beta$, we can 
thus apply our analysis to the WBB interface if we approximate the interface to be planar and stationary, which is reasonable as the radius of curvature of 
WBBs are much greater than the interface scale of relevance. We must also assume that the global magnetic field is primarily radial. 

The computational parameters used in our simulations and the real physics parameters typical in a WBB are listed in the first two columns of Tab.1. We 
choose the domain length to be $0.025\,pc$, which is about $1$ percent of the radius of the actual WBB. Tab.1 shows that by choosing the proper scaling, 
our simulation fits well with the data observed in a typical WBB. Therefore, the conclusions we draw by analysing the simulation results and the analytical 
expressions, especially Eq.(19), can be helpful in analyzing WBB evolution. 

\section{Conclusion}

We have investigated the problem of heat transfer in regions of initially arbitrarily  tangled magnetic fields in laminar high $\beta$ MHD flows  using simulation 
results of ASTROBEAR code with anisotropic heat conduction. Three conclusions stand out:

 (1) Hot and cold regions initially separated by a tangled field region with locally confined field loops may still evolve to incur heat transfer. The local redistribution 
of fluid elements bend the field lines and lead to magnetic reconnection that can eventually connect the hot and cold regions on the two sides. (2) The 
temperature gradient through such a penetrated tangeld field region tends to reach a steady state that depends on the energy difference between the hot 
and cold reservoirs on the two ends.  (3)  Eq.(19), a measure of the initial  field tangle, is a good predictor of the ultimate   heat transfer efficiencies across 
the interface for a wide range of $R$.

A basic limitation of our simulations  is that they are 2-D.  A 3-D version of this study would be of interest as the field would then have finite scales in the 
third dimension possibly allowing channels for heat transfer excluded in 2-D. We have also not considered the effects of cooling in our simulations.

Future directions of analysis could include a multi-mode study, which investigates the effect of the spatial spectrum of the magnetic field distribution on the heat 
transfer efficiency. When there are multiple modes or a spectrum is continuous, it would be useful to predict how the efficiency would depend on the spectrum. 
In this context, a more detailed comparison of heat transfer in initially laminar versus initially turbulent systems would be of interest.

\section{acknowledgements}
Financial support for this project was provided by the Space Telescope
Science Institute grants HST-AR-11251.01-A and HST-AR-12128.01-A;
by the National Science Foundation under award AST-0807363 and NSF PHY0903797; by the Department of Energy under award DE-SC0001063; and by Cornell University 
grant 41843-7012. We wish to thank Jonathan Carroll, Kris Yirak and Brandon Shroyer for useful discussions.

\appendix{\textbf{Appendix: Code Test}}
\paragraph{} The MHD solver and the linear thermal diffusion solver are verified by well-known tests such as the field loop convection problem and the Guassian diffusion 
problem separately. As a comprehensive test that involves both MHD and thermal diffusion, we use the magneto-thermal instability (MTI) problem to test the accuracy of the 
ASTROBEAR code with anisotropic heat conduction (\citet{par05},\citet{cun09}). The problem involves setting up a 2-D temperature profile with uniform gravity pointing on the 
y direction. The domain is square with length of $0.1$ in compuational units. The temperature and density profiles are:

\begin{equation}
T = T_0\,(1-y/y_0)
\end{equation}
\begin{equation}
\rho = \rho_0\,(1-y/y_0)^2
\end{equation}

\begin{figure}
\begin{center}
\includegraphics[width=\columnwidth]{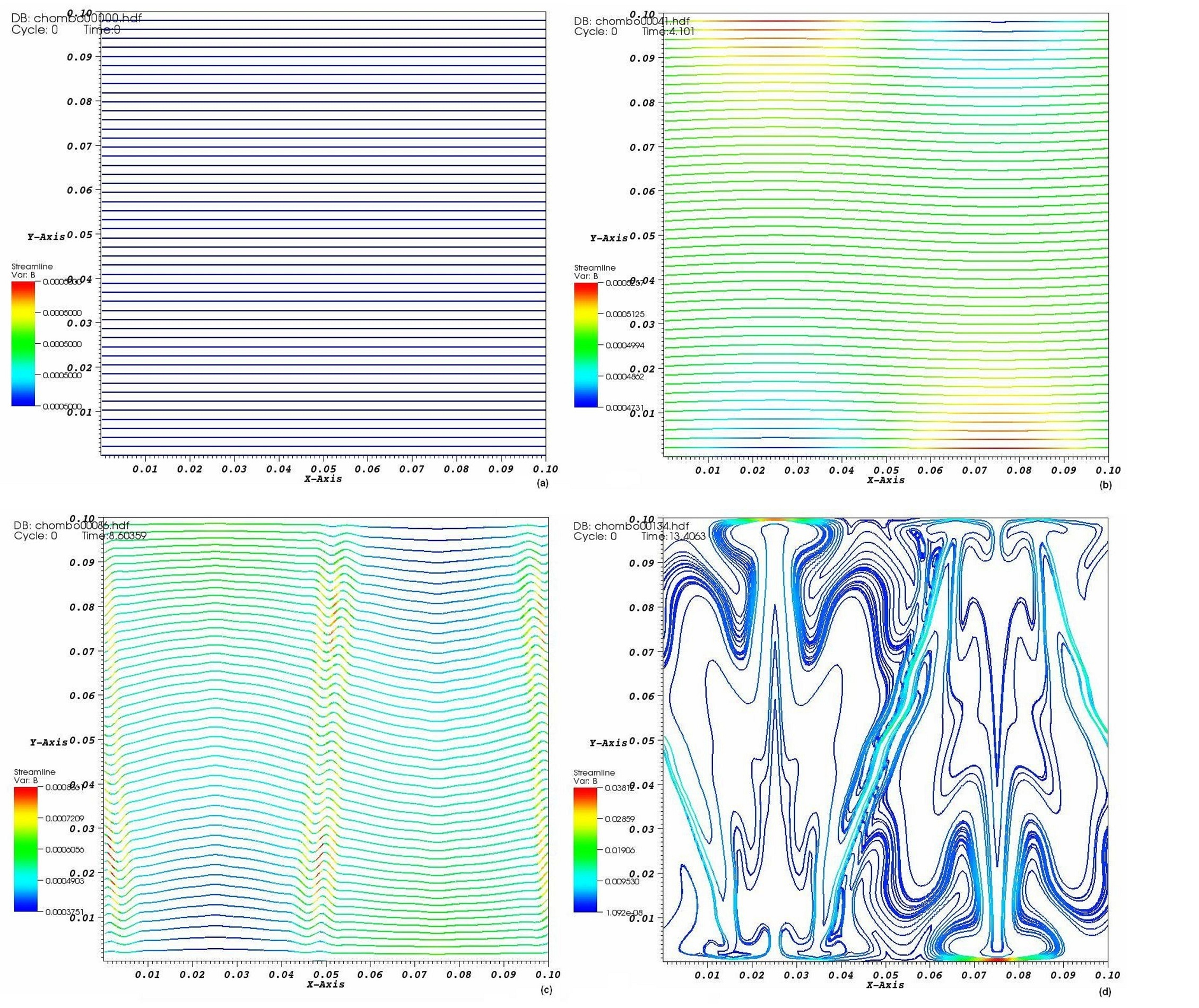}
\caption{Field line evolution of magneto-thermal instability. (a): initial state. (b): $t=75\tau_s$. (c)$t=150\tau_s$. (d):$t=250\tau_s$.}
\label{fig11}
\end{center}
\end{figure}

\placefigure{fig11}

\begin{figure}
\begin{center}
\includegraphics[width=.6\columnwidth]{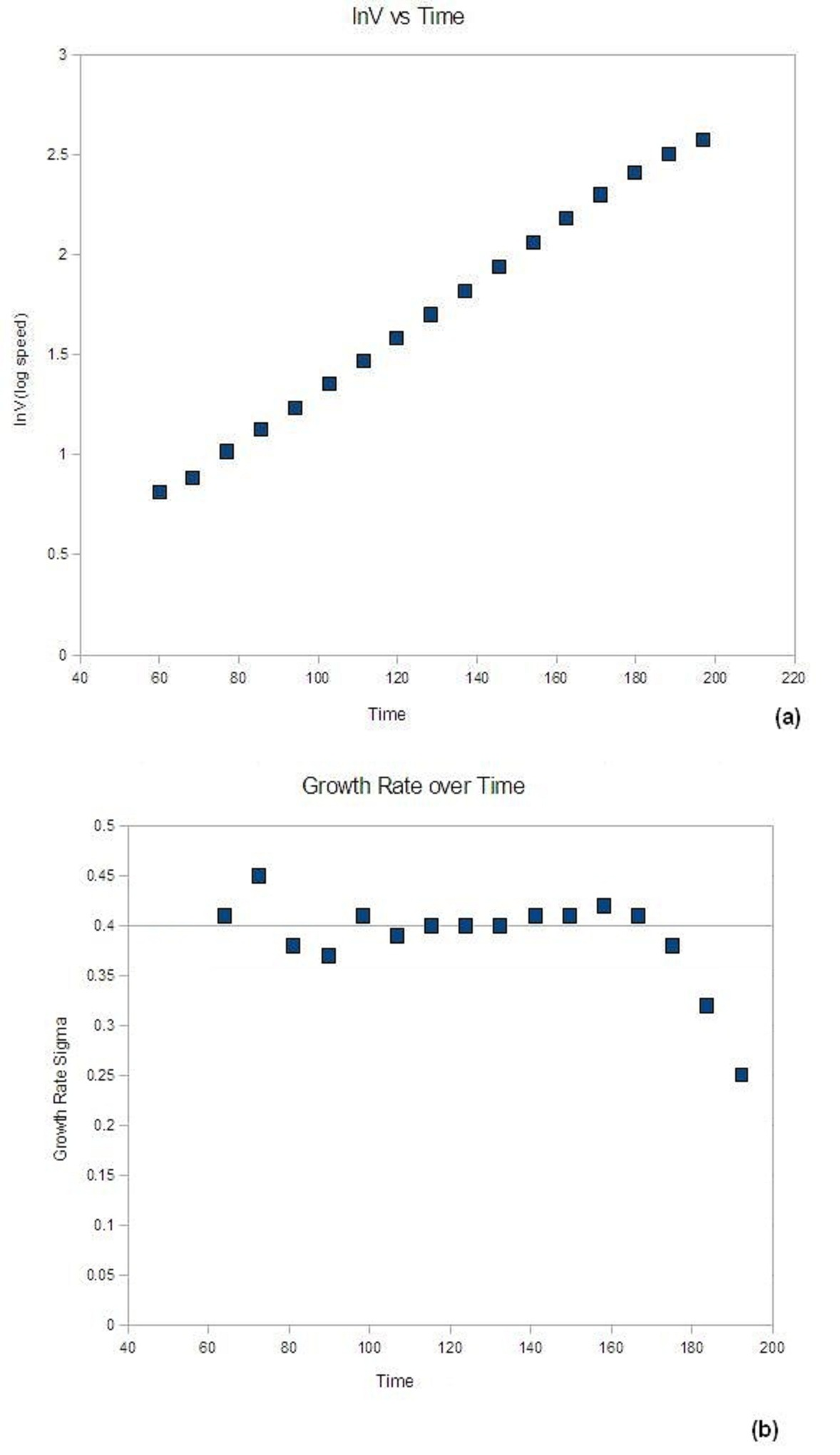}
\caption{(a): $\ln\,v_y$ against evolution time in $\tau_s$. (b): calculated growth rate against evolution time in computational units. Initially the growth rate is stable 
around the theoretical value $0.4$ and then decreases sharply after $t=200$, which indicates the evolution has entered the nonlinear regime.}
\label{fig12}
\end{center}
\end{figure}

\placefigure{fig12}

\begin{figure}
\begin{center}
\includegraphics[width=.8\columnwidth]{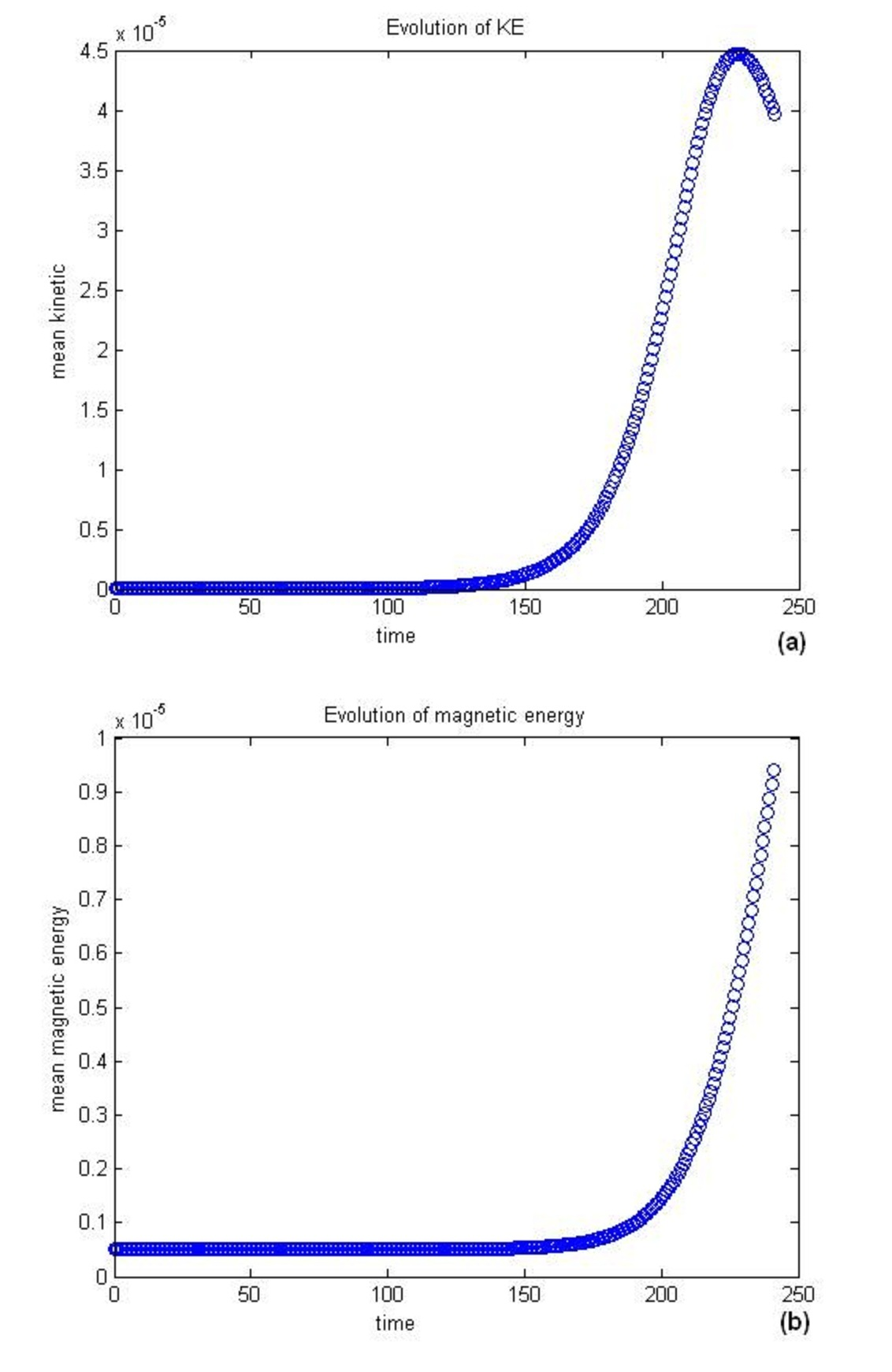}
\caption{(a): evolution of mean kinetic energy. (b): evolution of mean magnetic energy.}
\label{fig13}
\end{center}
\end{figure}

\placefigure{fig13}

with $y_0=3$. The pressure profile is set up so that a hydrostatic balance may be achieved with uniform gravity with gravitational acceleration $g = 1$ in computational units. 
We also set $T_0=1$ and $\rho_0=1$ in computational units. There is a uniform magnetic field on the x direction with field strength $B_0=1.0\times10^{-3}$ in computational units. 
The anisotropic heat conductivity is set to be $\kappa=1\times10^{-4}$ in computational units. We use the pressure equilibrium condition for the top and bottom boundaries, that is, 
the pressure in the ghost cells are set so that its gradient balances the gravitational force. On  the x direction, we use the periodic boundary condition.

Initially, the domain is in pressure equilibrium. We then seed a small velocity perturbation:

\begin{equation}
v_{per} = v_0\,\sin(n\,\pi\,x/\lambda)
\end{equation}

with $v_0=1\times10^{-6}$ and $\lambda=0.5$. This perturbation will cause the fluid elements to have a tiny oscillation on y axis as well as the field lines. Once the field lines are 
slightly bent, they open up channels for heat to transfer on the y direction thus allowing the heat on the lower half of the domain to flow to the upper half. It can be shown that 
this process has a positive feedback so that once the heat exchange happens, more channels will be openned up for heat conduction. Therefore this process forms an instability 
whose growth rate can be verified according to the linear instability growth theory. We use $\tau_s$ to denote the sound crossing time for the initial state. Figure~\ref{fig11} shows the time 
evolution of the field lines at various stages in our MTI simulation.

We study the MTI growth rate by considering the acceleration of the fluid elements. The mean speed on the y direction for the fluid should follow the exponential growth:

\begin{equation} 
v_y = v_{per}e^{\gamma\,t}
\end{equation}

where $v_{per}$ is the strength of the initial velocity perturbation applied, $\gamma$ denotes the growth rate in the linear regime. We obtain the growth rate $\gamma$ by plotting 
$\ln\,v_y$ against the evolution time and then measuring the local slope through a certain time span. The $\ln\,v_y$ vs $t$ curve is plotted in Figure~\ref{fig12}(a), which shows a nice linear 
relation. We plot the growth rate against evolution time. It should be stable around the theoretical value $0.4$ initially and then decrease sharply due to the nonlinear effect. 
Figure~\ref{fig12}(b). shows that the simulation meets our expectation fairly well.

We also look at the energy evolution in the linear regime. The mean kinetic energy should first stay stable and then enter into an exponential growing phase until it hits a cap at 
around $t=200$ which denotes the starting of the nonlinear phase. The evolution of magnetic energy should follow similar pattern as to the kinetic energy evolution, but lagged 
behind. In Figure~\ref{fig13}, we plot the time evolution of the mean kinetic and magnetic energy evolutions. The results confirms the physical intuition quite well.

\clearpage

\clearpage

\begin{table}
\caption{Scaling of Simulation Parameters}
\label{tab01}
\begin{tabular}{|c|c|c|}
\tableline
Variables & Computional Units & WBB \\
Number Density & $1$ & $1\,cm^{-3}$ \\
Temperature & $100$ & $1\,kev$ \\
Domain Length & $0.1$ & $0.025\,pc$ \\
Local Field Strength  & $10^{-3}$ & $2^{-8}\,Gauss$ \\
Global Field Strength  & $10^{-4}$ & $2^{-9}\,Gauss$ \\
Evolution Time  & $1.2$ & $12,000\,yrs$ \\
Heat Conductivity  & $10^{-2}$ & $2\times10^{-18}\,cm\,s\,g^{-1}\,K^{-2.5}$ \\
\tableline
\end{tabular}\\
\end{table}

\end{document}